\documentclass[twocolumn,superscriptaddress,showpacs,prl]{revtex4}
\usepackage{epsfig}
\usepackage{bm}

\newcommand{\rc}{\vec{r}_{\rm c}}
\newcommand{\kc}{\vec{k}_{\rm c}}
\newcommand{\rcdot}{\dot{\vec{r}}_{\rm c}}
\newcommand{\kcdot}{\dot{\vec{k}}_{\rm c}}
\newcommand{\zc}{z_{\rm c}}
\newcommand{\uzc}{\tilde{U}_{\zc}}

\newcommand{\prsl}[3]{Proc. R. Soc. A \textbf{#1}, {#2} (#3)}
\newcommand{\acta}[3]{Acta Crystallogr. \textbf{#1}, {#2} (#3)}
\newcommand{\annph}[3]{Ann. Phys. \textbf{#1}, {#2} (#3)}

\newcommand{\scien}[3]{Science \textbf{#1}, {#2} (#3)}
\newcommand{\phrb}[3]{Phys. Rev. B \textbf{#1}, {#2} (#3)}

\newcommand{\ibid}[3]{\textit{ibid}. \textbf{#1}, {#2} (#3)}
\newcommand{\phrl}[3]{Phys. Rev. Lett. \textbf{#1}, {#2} (#3)}
\newcommand{\naturemat}[3]{Nat. Mater. 
\textbf{#1}, {#2} (#3)}

\newcommand{\philmag}[3]{Philos. Mag. \textbf{#1}, {#2} (#3)}

\newcommand{\jpsj}[3]{J. Phys. Soc. Jpn. \textbf{#1}, {#2} (#3)}

\newcommand{\nucl}[3]{Nucl. Phys. B \textbf{#1}, {#2} (#3)}
\newcommand{\zphys}[3]{Z. Physik \textbf{#1}, {#2} (#3)}
\newcommand{\be}{\begin{eqnarray*}}
\newcommand{\en}{\end{eqnarray*}}

\begin{document}

\title{Dynamical Diffraction Theory for Wave Packet Propagation 
in Deformed Crystals}

\author{Kei Sawada}
\email{sawada@appi.t.u-tokyo.ac.jp}
\affiliation{Department of Applied Physics, University of Tokyo, 
7-3-1, Hongo, Bunkyo-ku, Tokyo 113-8656, Japan}

\author{Shuichi Murakami}
\affiliation{Department of Applied Physics, University of Tokyo, 
7-3-1, Hongo, Bunkyo-ku, Tokyo 113-8656, Japan}

\author{Naoto Nagaosa}
\affiliation{Department of Applied Physics, University of Tokyo, 
7-3-1, Hongo, Bunkyo-ku, Tokyo 113-8656, Japan}
\affiliation{Correlated Electron Research Center (CERC), 
National Institute of Advanced Industrial Science and Technology (AIST), 
Tsukuba Central 4, Tsukuba 305-8562, Japan}
\affiliation{CREST, Japan Science and Technology Agency (JST), Japan}

\date{\today}

\begin{abstract}
We develop a theory for the trajectory of an x ray 
in the presence of a crystal deformation. 
A set of equations of motion for an x-ray wave packet 
including the dynamical diffraction is derived,
taking into account the Berry phase 
as a correction to geometrical optics. 
The trajectory of the wave packet has a shift of the center position 
due to a crystal deformation. 
Remarkably, 
in the vicinity of the Bragg condition, 
the shift is enhanced by a factor $\frac{\omega}{\Delta \omega}$ 
($\omega$: frequency of an x ray, 
$\Delta\omega$: gap frequency induced by the Bragg reflection). 
Comparison with the conventional dynamical
diffraction theory is also made.

\end{abstract}

\maketitle


X-ray diffraction has been used as a powerful tool 
for analyzing crystal structures.
Conventional analysis of x-ray diffraction is based on 
a kinematical theory, namely the first-order Born approximation, 
and has played an important role in many fields 
beyond physics such as medical science. 
A theory incorporating the 
multiple-scattering process of x ray is called the 
dynamical diffraction theory \cite{authier}, 
pioneered by Ewald \cite{ewald} and Darwin \cite{darwin}.
In Darwin's theory for a perfect crystal, 
higher-order diffraction is taken into account by 
superposing the diffracted waves. 
A perfect crystal, however, suffers from 
an extinction effect;  
the intensity of the first-order diffracted wave is 
reduced by the higher-order waves. 
In order to avoid this difficulty, 
mosaicness is introduced into a crystal to suppress 
the contributions of the higher order diffracted waves 
by random interferences.
Thus, a perfect crystal is not preferable 
for the structural analysis, and 
applicability of the dynamical theory has been limited. 
However, recent technological advances for 
extremely intense source of x ray such as 
a free-electron laser will open up new fields of 
research associated with the dynamical theory.

On the other hand, 
for a deformed crystal, a dynamical theory was 
first developed dozens of years ago \cite{takagi, taupin, kato}.
In Refs. \cite{takagi, taupin}, 
eigenequations for the Bragg reflected beams 
are translated into differential equations. 
In general, 
it is almost impossible to analytically 
solve such equations, 
and it is not so easy to extract a clear 
physical picture from the numerical solutions.
We note a corresponding theory 
for electronic systems with dislocations 
\cite{howie, kawamura}.

These diffraction theories assume incident waves 
to be plane waves or spherical waves, and 
concern the propagation of their wave fronts. 
On the other hand, one can ask what happens 
if the incident wave is 
confined in the transverse direction, i.e., 
a narrow beam whose trajectory is well-defined. 
Naively one might expect that the trajectory of such a beam 
is always perpendicular to the wave front, which is 
expected in a wave propagation from Fermat's principle. 
In this Letter, 
we reveal an anomalous feature of such electromagnetic beams 
beyond this naive expectation. 
We derive equations of motion for an x-ray wave packet
in the presence of a crystal deformation.
These equations of motion incorporate 
the dynamical effect of a multiple scattering, 
and a geometrical phase associated with the wave dynamics,
i.e., a Berry phase \cite{Berry}, is taken into account in a natural way.
We find that such a Berry phase in deformed crystals 
gives rise to a shift of the center position of an x-ray 
wave packet, and enhances the shift 
by several orders of magnitude 
in the vicinity of the Bragg condition. 
Such a gigantic enhancement is physically 
related to a small group velocity 
near the Bragg condition.

A wave packet has a finite width 
both in real and momentum spaces due to 
the uncertainty principle. 
Such a wave packet feels ``curvatures" 
in real and/or momentum spaces. 
A Berry phase describes these curvatures in 
real and momentum spaces, 
and plays a nontrivial role in the wave dynamics 
in constrained systems. 
For electromagnetic waves, 
such a constraint is naturally realized; 
electromagnetic fields are always constrained to be 
perpendicular to the propagation direction 
described by the wave vector ${\vec k}$. 
The Berry phase for light appears 
for example in a twisted optical fiber in which   
the trajectory of the wave vector $\vec{k}$ makes a closed loop. 
In this case, the polarization plane rotates during propagation, 
and the rotation angle is represented by a Berry phase \cite{chiao}. 
Another kind of constraint is realized in periodic systems, 
where the dispersion relation has a band structure. 
When a physical state is constrained on each band, 
the Bloch function is associated with 
a Berry phase in momentum ($k$) space. 
Namely, $k$ space is curved for 
each band of the Bloch wave function,
which influences the wave dynamics. 
Onoda \textit{et al.} \cite{onoda} 
constructed a generalized geometrical optics 
including a Berry phase of light, 
and derived a set of equations of motion 
for an optical wave packet. 
The Berry phase in $k$ space gives rise to 
a transverse displacement of a wave packet, 
leading to the optical Hall effect.
They were mainly concerned with 
photonic crystals for visible light, 
and investigated how to design the 
photonic band structure \textit{in $k$ space}. 
On the other hand, this Letter focuses 
on the x-ray diffraction; 
the main interest here is how to probe a 
spatial deformation \textit{in real space}. 
Therefore the relevant Berry phase is distinct from that
discussed in \cite{onoda}.

\begin{figure}[tbp]
  \centerline{
    \epsfxsize=8.5cm
    \epsfbox{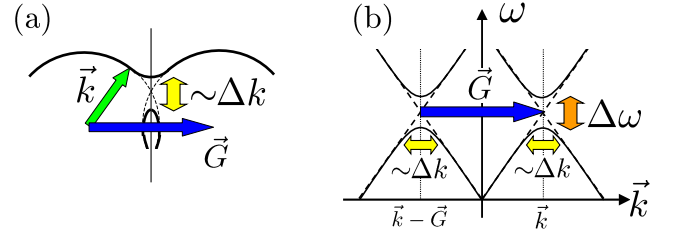}
  }
  \caption{(a) Constant-frequency surface 
and corresponding wave vector and reciprocal vector 
$\vec{k}$ and $\vec{G}$, respectively. 
(b) Dispersion relation. At the Bragg condition, 
$\vec{k}$ and $\omega$ have gaps 
$\Delta k$ and $\Delta\omega$.}
 \label{band}
\end{figure}

Let us introduce a six-dimensional 
electromagnetic field vector; 
\be
\vec{\Gamma}_{\vec{k}}(\vec{r})
=\frac{1}{\sqrt{\mathstrut 2}}
\left(
\begin{array}{c}
\sqrt{\varepsilon(\vec{r})}\vec{E}_{\vec{k}}(\vec{r}) \\
\sqrt{\mu(\vec{r})}\vec{H}_{\vec{k}}(\vec{r})
\end{array}
\right)
\equiv
\frac{1}{\sqrt{\mathstrut 2}}
\left(
\begin{array}{c}
\vec{F}^E_{\vec{k}}(\vec{r}) \\
\vec{F}^H_{\vec{k}}(\vec{r})
\end{array}
\right),
\en
where $\varepsilon$ and $\mu$ are 
the dielectric constant and the 
magnetic permeability, respectively. 
Hereafter we put $\mu=\hbar =1$. 
The wave equations for $\vec{F}^{E, H}$ are 
\be
\frac{1}{\sqrt{\mathstrut \varepsilon(\vec{r})}}\nabla \times 
\Bigl[ \nabla \times 
\frac{1}{\sqrt{\mathstrut \varepsilon(\vec{r})}}
\vec{F}^E(\vec{r}) \Bigr]
&=&\frac{\omega^2}{c^2}\vec{F}^E(\vec{r}),\\
\nabla \times 
\Bigl[ \frac{1}{\varepsilon(\vec{r})}
\nabla \times \vec{F}^H(\vec{r}) \Bigr]
&=&\frac{\omega^2}{c^2}\vec{F}^H(\vec{r}), 
\en
which are Hermitian. 
In these wave equations, we assume  
a periodic dielectric function in a crystal with respect to 
a primitive vector $\vec{a}$: 
$\varepsilon(\vec{r}+\vec{a})=\varepsilon(\vec{r})$. 
In the dynamical theory, 
the eigenmodes of an electromagnetic wave 
are Bloch functions under the periodic condition. 
They are written in the six-dimensional representation as  
$ 
\vec{\Gamma}_{\vec{k}\zc}(\vec{r})
=e^{i\vec{k}\cdot\vec{r}}\tilde{U}_{\vec{k}\zc}(\vec{r})$, 
where $\tilde{U}_{\vec{k}\zc}(\vec{r})$ is 
a periodic function satisfying 
$\tilde{U}_{\vec{k}\zc}(\vec{r}) 
=\tilde{U}_{\vec{k}\zc}(\vec{r}+\vec{a})$, 
and $\zc$ represents the polarization state.
Figure~\ref{band}(a) shows 
the constant-frequency contour in $k$ space, 
where the wave vector has a gap at the Bragg condition. 
Such a gap in $\vec{k}$ 
corresponds to an energy gap at the Bragg condition 
$|\vec{k}|=|\vec{k}-\vec{G}|$, as shown in Fig.~\ref{band}(b). 
In the energy gap, the wave number 
becomes pure imaginary and no x ray can propagate
through a crystal. 
Due to this gap,  
the dispersion relation for the Bloch function 
has a band structure, 
and the Bloch functions are specified by 
a band index $n$; 
$\vec{\Gamma}_{n\vec{k}\zc}(\vec{r})$. 
Using the Bloch functions, 
we construct a wave packet 
whose distribution function  
in $k$ space has a sharp peak 
around the center of the momentum $\kc$. 
The wave packet has a peak 
around the center position $\vec{r}=\rc$ in real ($r$) space . 
We consider slowly varying 
perturbations which do not induce  
the interband transitions, 
and keep only the diagonal matrix elements
with respect to the band index $n$. 
The distribution in $k$ space gives 
an interference between components of different wave vectors, 
which can be expressed in terms of 
the Berry curvature in $k$ space 
associated with each band. 
In addition, 
we introduce a crystal deformation whose spatial variation 
is small within the width of the wave packet. 
Such a slowly varying deformation  
leads to the Berry connection in $r$ space, 
as we will show later.

Let us derive equations of motion for such a wave packet. 
An effective Lagrangian for a wave packet is 
calculated in a similar manner as in \cite{niu, onoda} 
to be
\be
L(\rc, \kc, \dot{\rc}, \dot{\kc}, \zc, \dot{\zc})
=-\omega_n + \kc\cdot\rcdot+i\dot{\rc}\cdot 
\langle \tilde{U}_{\zc} |\nabla_{\rc}\tilde{U}_{\zc} \rangle \\
+i\dot{\kc}\cdot 
\langle \tilde{U}_{\zc} |\nabla_{\kc}\tilde{U}_{\zc} \rangle
+i\langle \tilde{U}_{\zc} |\partial_t\tilde{U}_{\zc} \rangle. 
\en
The first two terms on the right-hand side describe  
the wave front propagation, which exist even for 
plane waves or spherical waves. 
The last three terms represent 
the additional physics which appears 
only in the wave packet dynamics, 
leading to Berry curvatures. 
This effective Lagrangian is 
analogous to that for electronic systems \cite{niu, shindou}.
Variational derivative of the Lagrangian gives 
the following equations of motion; 
\begin{eqnarray}
\dot{\kc}&=&
-\frac{\partial \omega_n}{\partial \rc}
+\Omega_{\vec{r}\vec{k}}\cdot \kcdot
+\Omega_{\vec{r}\vec{r}}\cdot \rcdot
-\Omega_{t\vec{r}},\\
\dot{\rc}&=&
\frac{\partial \omega_n}{\partial \kc}
-\Omega_{\vec{k}\vec{k}}\cdot \kcdot
-\Omega_{\vec{k}\vec{r}}\cdot \rcdot
+\Omega_{t\vec{k}},\\
|\dot{\zc} )&=&
i\rcdot\cdot\vec{{\cal A}}_{\rm r}|\zc )
+i\kcdot\cdot\vec{{\cal A}}_{\rm k}|\zc ),
\end{eqnarray}
where  
\be
( \Omega_{\vec{r}\vec{k}})_{\alpha \beta}
\equiv
\Omega_{r_\alpha k_\beta}
&\equiv&
\langle \partial_{r_\alpha}\uzc | 
i\partial_{k_\beta}\uzc \rangle
- \langle \partial_{k_\beta}\uzc | 
i\partial_{r_\alpha}\uzc \rangle , \\
\bigl[
\vec{{\cal A}}_{\rm r}
\bigr]_{\lambda\lambda'} &\equiv&
i\langle \tilde{U}_{n\lambda \kc \rc} 
|\nabla_{\rc} 
\tilde{U}_{n\lambda' \kc \rc}
\rangle, \\
\bigl[
\vec{{\cal A}}_{\rm k}
\bigr]_{\lambda\lambda'} &\equiv&
i\langle \tilde{U}_{n\lambda \kc \rc} 
|\nabla_{\kc} 
\tilde{U}_{n\lambda' \kc \rc}
\rangle,
\en
$|\zc)$ has two components representing the 
polarization states, and $\lambda$ is a polarization index. 
The vectors $\vec{\cal A}_{\rm r, k}$, called 
Berry connections in mathematics, 
are ``vector potentials" in such a sense that 
their rotation gives the Berry curvatures. 
Each term in the above equations of motion can be interpreted 
as follows. 
The first terms on the right-hand side of Eqs.~(1) and (2) 
reproduce geometrical optics. 
The second term in Eq.~(1) represents 
effects of the crystal deformation, while
the third one corresponds to a Lorentz force 
which is zero in a nonmagnetic system. 
The second term in Eq.~(2) 
represents the anomalous velocity of light \cite{niu, murakami}, 
giving rise to an optical Hall effect \cite{onoda}.
The third term represents the displacement of 
a ray due to a crystal deformation. 
Equation~(3) describes the time evolution of the polarizations. 
Since we consider a nonmagnetic system,  
$\Omega_{\vec{r}\vec{r}}=0$. 
We also assume that the perturbations are independent of time; 
$\Omega_{t\vec{r}}=\Omega_{t\vec{k}}=0$.

\begin{figure}[htbp]
  \centerline{
    \epsfxsize=7.5cm
    \epsfbox{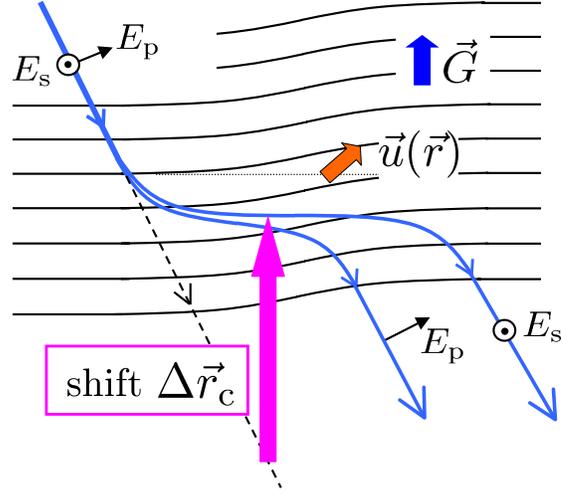}
  }
  \caption{Schematic illustration of the deformed crystal 
with an atomic displacement vector $\vec{u}(\vec{r})$, 
The trajectory of the center position of an x-ray wave packet 
is represented by the oriented curve. 
The amount of the shift can be as large as 
$\Delta \rc \sim 0.1$ mm.
The broken line with an arrow represents the trajectory 
without the deformation. 
$E_{\rm s}$ and $E_{\rm p}$ denote the electric fields for 
s and p polarizations, respectively.}
 \label{displacement}
\end{figure}

Consider a system with a deformation, 
as shown in Fig.~\ref{displacement}, 
where 
the spatial variation of the deformation is small within  
the width of the wave packet. 
We assume that the wave packet 
has a sharp peak in $k$ space around the center $\kc$, 
with its width much smaller than $|\vec{G}|$. 
This implies that 
the peak in $r$ space is much wider than 
the lattice constant.
In this situation, 
the eigenfunction of an x ray is modified as
$\vec{\Gamma}(\vec{r})
\rightarrow 
\vec{\Gamma}(\vec{r}-\vec{u}(\rc))$, 
where $\vec{u}(\vec{r})$ is a continuous function 
representing the atomic displacement \cite{atom}. 
In an equation of motion for $\kc$, 
we assume $\frac{\partial \omega_n}{\partial \rc}=0$ 
as the first approximation, 
resulting in $\kcdot=0$. 
Hence in the present level of approximation, 
$\Omega_{\vec{k}\vec{k}}$ does not affect the result. 
Thus we have only to consider the terms including 
$\Omega_{\vec{r}\vec{k}}$ and $\Omega_{\vec{k}\vec{r}}$. 
The Berry connection between the neighboring $\rc$ points 
is calculated as 
\be
\Bigl\langle \tilde{U}_{\zc} \Bigl|
i\frac{\partial\tilde{U}_{\zc}}{\partial \rc}
\Bigr\rangle
=
\Bigl\langle \tilde{U}_{\zc}(\vec{r}-\vec{u}(\rc)) \Bigl|
i\frac{\partial}{\partial \rc}\tilde{U}_{\zc}(\vec{r}-\vec{u}(\rc))
\Bigr\rangle\\
=
-\frac{\partial u_\beta}{\partial \rc} 
\Bigl\langle \tilde{U}_{\zc}(\vec{r})\Bigl| 
i\frac{\partial}{\partial r_\beta} 
\tilde{U}_{\zc}(\vec{r}) 
\Bigr\rangle 
=
-\frac{\partial u_\beta}{\partial \rc}
{\cal A}_{{\rm r}\beta}(\kc),
\en
where $\vec{\cal A}_{\rm r}(\kc)$ is the Berry connection 
for a fixed polarization state 
in the absence of the deformation, 
and repeated subscripts are summed over. 
Therefore, the Berry curvature term 
in the equation for $\rc$ is 
\be
(\Omega_{\vec{k}\vec{r}}\cdot \rcdot)_\alpha
=-\frac{\partial u_\beta}{\partial r_{{\rm c}\gamma}}
\frac{\partial {\cal A}_{{\rm r}\beta}}{\partial k_{{\rm c}\alpha}}
\dot{r}_{{\rm c}\gamma}. 
\en
This is a product of the two factors. 
One is $\nabla_{\rc}\vec{u}(\rc)$ 
due to the deformation, and 
the other is $\nabla_{\kc}\vec{\cal A}_{\rm r}(\kc)$ 
originated from a band structure. 
Therefore, one can see that 
the Berry curvature $\Omega_{\vec{k}\vec{r}}$ represents 
an interplay between 
the curvatures in $k$  and $r$ spaces. 
Using these expressions, we can integrate the 
equation and obtain the displacement of an x ray due to 
the crystal deformation as 
\be
r_{{\rm c}\alpha}-r_{0\alpha}
&=&v_{{\rm g}\alpha}t
-\!\int 
(\Omega_{\vec{k}\vec{r}}\cdot \rcdot)_\alpha dt
=v_{{\rm g}\alpha}t+
\frac{\partial {\cal A}_{{\rm r}\beta}}{\partial k_{{\rm c}\alpha}}
\int \!\!du_\beta \\
&=&v_{{\rm g}\alpha}t+
\frac{\partial \vec{\cal A}_{\rm r}}{\partial k_{{\rm c}\alpha}}
\cdot \Bigl[ 
\vec{u}(\rc)-\vec{u}(\vec{r}_0)
\Bigr],  
\en
where $\vec{v}_{\rm g}\equiv 
\frac{\partial \omega}{\partial \kc}$ 
is a group velocity, 
and $\vec{r}_0 \equiv \rc(t=0)$. 
We show below that the factor 
$\frac{\partial {\cal A}_\beta}{\partial k_{{\rm c}\alpha}}$ 
enhances the displacement by several orders of magnitude 
in the vicinity of the Bragg condition.  
We assume that the wave vector is in the vicinity of 
the Bragg condition,  
$|\frac{\vec{k}^2-(\vec{k}-\vec{G})^2}{2\vec{k}^2}|
\ll \frac{\Delta\omega}{\omega}
\sim \frac{\Delta k}{k}$, 
where $\Delta \omega$ and $\Delta k$ are gaps in 
$\omega$ and $\vec{k}$, respectively. 
The Bloch function is expressed as 
$\vec{\Gamma}(\vec{r})
\simeq
\vec{\Gamma}_0e^{i\vec{k}\cdot\vec{r}}
+\vec{\Gamma}_1e^{i(\vec{k}-\vec{G})\cdot\vec{r}}$ 
in two-wave approximation. 
From this wave function, the enhancement factor 
$\frac{\partial {\cal A}_{{\rm r}\beta}}{\partial k_{{\rm c}\alpha}}$ 
is calculated as 
\be 
\frac{\partial {\cal A}_{{\rm r}\beta}}{\partial k_{{\rm c}\alpha}} 
\Bigl|_{|\vec{k}|\simeq |\vec{k}-\vec{G}|} 
&\simeq& 
\pm \frac{1}{2} 
\frac{G_\alpha G_\beta}{|\vec{k}|^2} \frac{\omega}{\Delta \omega}, 
\en 
where $+$ and $-$ denote 
the lower and the upper bands, respectively \cite{note}. 
Therefore, the solution to the equations of motion 
near the Bragg condition is 
\begin{equation} 
\rc -\vec{r}_0 
\ \simeq \ \vec{v}_{\rm g}t\pm 
\vec{G}\Bigl[ 
\vec{G}\cdot \Bigl( 
\vec{u}(\rc)-\vec{u}(\vec{r}_0)\Bigr) 
\Bigr] 
\frac{\omega}{2\Delta \omega} 
\frac{1}{|\vec{k}|^2}. 
\end{equation}
This formula is the central result of this Letter. 
The direction for the shift is parallel to $\vec{G}$,  
and the amount of the shift becomes maximum when 
$\vec{G}\parallel \vec{u}$. 
Such a shift is enhanced by 
a factor $\frac{\omega}{\Delta \omega}$. 
This factor differs for polarization states, 
s- and p-polarizations, which are defined 
relative to the plane
spanned by $\vec{k}$ and $\vec{k}-\vec{G}$.
The different size of the gap for each polarization state 
gives the different amount of the shift. 
When we inject an unpolarized beam, 
it splits into two beams with s- and p-polarizations, 
as shown in Fig.~\ref{displacement}.

Now we give a realistic estimate for the enhancement factor 
$\frac{\omega}{\Delta \omega}$ 
and the shift $\Delta \rc$ 
shown in Fig.~\ref{displacement}. 
For an x ray, the dielectric constant 
can be approximated as  
$\varepsilon(\vec{r})=1-\omega_{\rm p}^2/\omega^2$, 
where $\omega_{\rm p}$ is a plasma frequency, and typically    
$\omega_{\rm p}^2/\omega^2
 \sim 10^{-6}$ \cite{authier}. 
Hence the value of the enhancement factor for 
an x ray can be $\frac{\omega}{\Delta\omega}\sim 10^6$, 
which yields $\frac{\Delta k}{k}\sim 10^{-6}$. 
For a deformation of $|\vec{u}|\sim 0.1$ nm, 
the displacement can be 
$\Delta\rc \sim 0.1$ mm. 
It means that the deformation in an atomic scale 
gives rise to a macroscopic shift of the wave packet. 
Therefore, 
the factor $\frac{\omega}{\Delta \omega}$ plays the role 
of a ``lens", which gigantically magnifies 
the geometrical effect on the wave packet dynamics. 
Such a gigantic enhancement of the shift 
is a consequence of  
an energy gap and can be physically interpreted as follows. 
In the vicinity of the Bragg condition, 
we have a very small group velocity $\vec{v}_{\rm g}$. 
Such a slow wave packet has enough time to interact with electrons, 
resulting in a remarkably large optical effect. 
This kind of an enhancement at the edge of the energy gap 
is often studied in optical waves in  photonic crystals \cite{soljacic}. 
Photonic crystals, however, have a large energy gap, 
and hence the enhancement is much smaller.

We note that 
the distribution of the wave packet in $k$ space  
is desired to be narrower than $\Delta k$. 
The components of wave numbers outside of this $\Delta k$ 
width do not show the enhanced shift.
To make the whole wave packet shift, 
an x ray with a wider spot size 
$l_{\rm r} \sim (\Delta k)^{-1} \sim 0.1$ mm is desirable,  
because its peak width in $k$-space 
is as small as $\Delta k \sim 10^{-6}|\vec{G}|$. 
The amount of the shift for such a beam is  
$\Delta\rc \sim l_{\rm r}$. 
For a typical beam whose spot size is 
$l_{\rm r}\sim 100$ nm, 
only a small portion of the wave packet is shifted, 
and rest propagates without shifts. 
It is because 
the width in $k$ space of this wave packet is 
larger than the gap $\Delta k$.  
It is then difficult to measure the shift.

Let us compare the present theory 
with the conventional diffraction theory. 
In the conventional kinematical theory, 
only an equation of motion for 
$\kc$ is concerned. 
The equation for $\kc$ describes 
the propagation of the wave front. 
The term $\frac{\partial \omega_n}{\partial \rc}$ 
can deflect the beam; 
its amount, however, is not enhanced by the dynamical effect. 
The conventional dynamical theory deals with 
a part of the wave front 
which propagates through a deformation, and 
calculate the multiple scatterings which result in refraction. 
Although a small part of the wave front has often been 
regarded as a wave packet, 
our theory reveals a novel physics of a wave packet, 
apart from wave front physics.
The anomalous enhancement of the shift  
of the wave packet in this Letter 
occurs because the wave packet is not a monochromatic wave, 
but a mixture of waves with slightly different wave numbers. 
An interference between these waves gives rise to the 
enhancement discussed in this Letter. 
This physics of a wave packet is absent in a plane wave 
or a spherical wave.

It should be warned that 
such a displacement is completely different from refraction. 
In refraction, the spot position on the screen is dependent on 
the distance between a sample and a screen, 
because the refraction is caused by the shift of $\kc$. 
On the other hand, the shift of $\rc$ obtained in this Letter 
is independent of the distance. 
Hence the shifts of $\kc$ and $\rc$ 
have different physical origins 
and are experimentally distinguishable.

In summary, 
we have developed a dynamical diffraction theory 
for a wave packet propagation in a deformed crystal. 
Equations of motion for a wave packet are derived. 
The central result is 
the shift of the trajectory given in Eq.~(4); 
the center of position of the wave packet is 
displaced by a crystal deformation parallel to the 
reciprocal lattice vector $\vec{G}$, 
and the amount of the shift has maximum 
when $\vec{G}\parallel \vec{u}(\vec{r})$. 
Remarkably, such a shift of the wave packet is 
gigantically enhanced 
by the factor $\frac{\omega}{\Delta \omega}$, 
which is $\sim 10^6$ for an x ray. 
Since the factor is 
different for each polarization state, 
the wave packets with no-polarization splits into 
two beams with independent polarizations. 
In contrast with the conventional dynamical theory 
which focuses on the wave front and 
requires numerically solving the Maxwell equations, 
our theory treats the narrow beam and 
provides an easy way to evaluate 
a displacement of the wave packet for a given deformation.
An x-ray wave packet with a good coherency is 
useful to measure a deformation or dislocation in a crystal.

This work is financially supported by 
Grants in Aid from the Ministry of Education, Culture, 
Sports, Science and Technology of Japan. 
K. S. is supported by the Japan Society for the 
Promotion of Science.


\begin{thebibliography}{99}
\bibitem{atom} The displacement of an atom located at $\vec{R}_i$ is $\vec{u}_i$. 
The continuous function $\vec{u}(\vec{r})$ satisfies 
$\vec{u}(\vec{R}_i+\vec{u}_i)=\vec{u}_i$.
\bibitem{note}This expression is obtained as follows. 
First, solve the wave equations for each polarization 
and explicitly calculate the electromagnetic field vectors $\vec{F}^{E, H}$ 
in the two-wave approximation. 
Second, calculate the Berry curvature 
$\langle \tilde{U}|\nabla_{\rc}\tilde{U}\rangle$, 
and take the $\kc$-derivative.
\end{thebibliography}
\end{document}